\documentclass[aps,pra,preprint]{revtex4-1}
\usepackage{graphicx,xcolor}
\usepackage{amsmath,amssymb}
\usepackage{dsfont}

\newcommand{\Hq}{\mathcal{H}}
\newcommand{\heff}{H_{\text{\tiny eff}}}

\newcommand{\Lhl}{\mathcal{H}_{\text{\tiny L}}}
\newcommand{\hl}{H_{\text{\tiny L}}}
\newcommand{\ul}{U_{\text{\tiny L}}}
\newcommand{\hsl}{\mathcal{H}_{\text{\tiny SL}}}
\newcommand{\Lhsl}{\mathcal{H}_{\text{\tiny SL}}}
\newcommand{\hdd}{\mathcal{H}_{\text{\tiny DD}}}
\newcommand{\Lhloc}{\mathcal{H}_{\text{\tiny loc}}}

\newcommand{\nn}{\nonumber}

\newcommand{\tr}{\text{Tr}}
\newcommand{\trl}{\text{Tr}_{\text{L}}}
\newcommand{\rh}{\rho_{s}}
\newcommand{\rl}{\rho_{L}^{eq}}

\newcommand{\etal}{\textit{and others }}
\newcommand{\sts}{\vert_{t\rightarrow\infty}}
\newcommand{\stoo}{\vert_{t\rightarrow 0}}

\newcommand{\ket}[1]{\vert #1 \rangle}

\newcommand{\comment}[1]{ }
\begin{document}
\title{Effects of dipolar coupling on an entanglement storage device}

\author{Saptarshi Saha}
\email{ss17rs021@iiserkol.ac.in}
\author{Rangeet Bhattacharyya}
\email{rangeet@iiserkol.ac.in}

\affiliation{ Department of Physical Sciences, 
Indian Institute of Science Education and Research Kolkata,
Mohanpur - 741 246, WB, India }
 
\date{\today}
\begin{abstract}

Quantum computation requires efficient long-term storage devices to preserve quantum states. An attractive
candidate for such storage devices is qubits connected to a common dissipative environment. The common
environment gives rise to persistent entanglements in these qubit systems. Hence these systems act
efficiently as a storage device of entanglement. However, the existence of a common environment often
requires the physical proximity of the qubits and hence results in direct dipolar coupling between the
qubits. In this work, we investigate effects of the secular and the nonsecular part of the dipolar coupling
on the environment-induced entanglement using a recently-proposed fluctuation-regulated quantum master
equation [A. Chakrabarti and R. Bhattacharyya, Phys.  Rev. A 97, 063837 (2018)]. We show that nonsecular
part of the dipolar coupling results in reduced entanglement and hence less efficiency of the storage
devices. We also discuss the properties of efficient storage that mitigates the detrimental effects of the
dipolar coupling on the stored entanglement.

\end{abstract}
\maketitle

\section{Introduction}\label{intro}

In quantum information processing, the generation and the preservation of entanglements are particularly
important operations. The preservation or storage of entanglements is challenging since all quantum systems are,
to some extent, coupled to their environments. The dissipation originating from the environment destroys
the coherence and entanglement in quantum systems. Naturally, a method of storage 
that takes into account the environment's dissipative nature and yet provides a long-lived entangled 
state is highly desired.



One of the earliest clues to solve this problem was found when
two decades ago, Cabrillo \etal and Plenio \etal separately reported the generation of spatially separated
entangled atoms \cite{cabrillo1999, plenio1999, beige2000}. While Cabrillo \etal created the entanglement by
simultaneous excitation of the atoms, Plenio \etal used a leaky cavity to couple the atoms. In this sense,
the later work is possibly the first use of a dissipative environment to generate entanglement in spatially
distant atoms. At the same time, Arnesen \etal investigated the entanglement in a Heisenberg 1D spin-chain,
where the coupling gives rise to the entanglement but not in the presence of a heat bath \cite{arnesen2001}.
Within a few months, several groups reported that a common dissipative environment gives rise to entangled
quantum systems \cite{schneider2002, kim2002, basharov2002, jacobczyk2002}. Subsequently, Braun showed that
the generation of the entangled states may not require a direct coupling between the quantum systems and can
persist for arbitrarily long times \cite{braun2002}. Later, Benatti \etal demonstrated in a series of works that
Markovian quantum master equations (QME) could be used to arrive at the persistent bath-induced
entanglements \cite{benatti2003, benatti2004, benatti2006a, benatti2006b}. They have also found that the
final entangled state might depend on the initial value of the entanglement \cite{benatti2006a,
choi2007}. Later a complete characterization of the entangled phase for a pair of resonant oscillators
connected to a common bath was provided by Paz \etal \cite{paz2008}. In the remaining parts of the
manuscript, we shall refer to this type of entanglement as the environment-induced entanglement
(EIE). 

When a pair of qubits are coupled to a single common dissipative environment, one of the eigenstates of the
Lindbladian is the singlet state density matrix. Naturally, a pair of qubits prepared in this state will
remain unaffected by the dissipator. This remains the basic working principle of the entanglement storage
solution based on a bipartite system in the common environment. However, the particular form of the
dissipator from the common environment is strongly sensitive to the distance between the qubits. In an
important work, Zell \etal showed that the entanglement generated by a heat bath is persistent only if the
distance between the quantum systems is small compared to the cutoff wavelength of the system-bath
interaction \cite{zell2009}. The physical proximity of the qubits indicates that the qubits could be dipolar
coupled.

The works of Zell, McCutcheon, and Jeske showed that the distance between the non-interacting qubit pair
must be small to have a persistent entanglement; otherwise, the entanglement decays exponentially with the
increasing distance between the qubits \cite{zell2009, mccutcheon2009, jeske2013}. The analysis relies on
the fact that the bath correlation decreases exponentially with a characteristic length scale for a simple
bosonic bath.  In general, one should also include a qubit pair's direct dipolar interaction and
system-environment coupling in the dynamics. After all, if one restricts the distance between the qubit pair
to keep those within the bath-correlation length, it is expected that a dipolar interaction between this
qubit pair would have a measurable effect. To keep track of various terms appearing in the dipolar
interaction, following standard practice, we divide them into secular and nonsecular parts, where the former
commutes with the Zeeman part of the qubit Hamiltonians \cite{duer2001}.

An important aspect of the dipolar coupling between the qubits is the contribution from the nonsecular part
of the dipolar Hamiltonians. We have shown recently that such terms significantly influence the lifetime of
coherences in the case of a time-periodic Hamiltonian \cite{saha2022}. Moreover, the seminal works of Bloembergen
\etal established long ago that the nonsecular parts of the dipolar interactions contributed to the
transition probabilities \cite{bpp1948}. Therefore, such terms should be included in analyzing two or higher
qubit networks.  Now despite having a very large volume of literature on this problem, the effect of the
dipolar coupling on persistent entanglements has not hitherto been investigated. So, in this work, we
specifically seek an answer to the following question: in the context of a dipolar coupled qubit pair in a
spatially-correlated and fluctuating environment, how does the dipolar interaction affect the persistent
entanglement? A particularly important aspect of this question is whether the higher-order effects of the
dipolar coupling play a role in these dynamics. We note that one needs an appropriate quantum master
equation to include the aforementioned effects in the equations of motion of the dipolar systems.

In the traditional form of the quantum
master equation, the system-environment coupling provides the dissipator, and other interactions, such as the
drive or the dipolar interactions, contribute to the first-order processes. The fluctuation-regulated
quantum master equation shows that other interactions, can have a Redfield-like second-order effect in the
dynamics in addition to the dissipator from the system-environment coupling \cite{chakrabarti2018b}. The
dissipator from the drive, which provides a non-Bloch decay, has also been verified by Chakrabarti \etal
\cite{chakrabarti2018a}. Later, Chanda \etal have shown that this drive-induced dissipation from FRQME
results in an optimal condition of the quantum gates \cite{chanda2020}. Moreover, Chatterjee \etal have
recently demonstrated that the FRQME could successfully explain the frequency-dependent non-linear behavior
of the light shifts and Bloch-Siegert shifts \cite{chatterjee2020b}. Recently, we have studied how such
effects give rise to the decoherences in a dipolar coupled systems \cite{saha2022}. 

So, in this manuscript, we systematically study the effect of these terms on quantum storage solutions.
We shall show that a direct dipolar coupling reduces the magnitude of the stored entanglement. 





We organize the manuscript in the following order.  The description of the system consisting of dipolar
coupled qubits, weakly interacting with a spatially correlated bosonic bath is given in section
\ref{hamsys}. The dynamical equation for the system using FRQME is introduced in terms of symmetric and
anti-symmetric observables in section \ref{equations}. The dynamical behaviors and the steady-state
solutions which have been analyzed for different parameter values, are described in section \ref{results}. In the same
section, we also demonstrate the generation of the entanglement and their dependence on the dipolar
interactions. Finally, we discuss the results, and their implications on the EIE-based
storage devices in section \ref{discussions}.  In appendix \ref{derive-frqme}, we provide a brief sketch of
the derivation of the recently-proposed FRQME. The complete set of the dynamical equations, including the
dipolar coupling and a spatially correlated environment is given in appendix \ref{alleqs}.

\section{The model}
\label{hamsys}

As a specific realization of a qubit, we choose a spin-1/2 particle placed in a static homogeneous magnetic
field. We consider two such particles coupled to an environment that is in thermal equilibrium at an
inverse temperature $\beta$. We assume that the spins are dipolar
coupled whose strength scales as $1/r^3$, where $r$ is assumed to be the distance between the two qubits
\cite{duer2001}. 


The full Hamiltonian of the system and the environment can be written as,
\begin{eqnarray}
\Hq =\Hq_{1}^{\circ}+\Hq_{2}^{\circ}+\Hq_{L}^{\circ}+\hsl +\hdd+\hl(t),
\end{eqnarray}
where,$\Hq_{n}^{\circ} = \frac{\omega_{\circ}}{2} \sigma_z^{n}$ is the Zeeman Hamiltonian of the $n^{\rm th}$ qubit, 
$\sigma_{\alpha}, \alpha \in {x,y,z}$ are Pauli matrices for spin-1/2, and $\omega_{\circ}$ is the Larmor
frequency. For simplicity, we assumed that the Larmor frequencies of the two spins are identical.
$\Hq_{L}^{\circ}$ is the free Hamiltonian of the environment and can be modeled as a large collection of
harmonic oscillators which mimics a spatially correlated bosonic bath, to which the spins are coupled.



The construction of the system-environment coupling is largely motivated by the works of McCutcheon \etal
and Jeske \etal. Following their works, we write the Hamiltonian of system-environment coupling as $\hsl =
\sum\limits^2_{n=1}(\sigma_{+}^{n} \mathcal{L}_{-} B(t) \eta(\vec{r}_n)  + \sigma_{-}^{n} \mathcal{L}_{+}
B^{\star}(t)\eta^{\star}(\vec{r}_n))$ where, $B(t)$ is the system-environment coupling strength between the spin and the
environment, $\eta(\vec{r}_n)$ is the spatial part of the field due to the environment at the position of
$n^{\rm th}$ spin, $\sigma_{\pm}^{n}$ are the raising and the lowering operators for $n^{\rm th}$ spin,
$\mathcal{L}_{+}$ and $\mathcal{L}_{-}$ are the raising and the lowering operators of the enviroment, which,
for simplicity, are assumed to be resonant
with the qubit pair. The correlation between the $\eta(\vec{r})$ decides
the nature of the environment. One may choose a suitable correlation to render the environment to be local
or common, as discussed in more detail in the next section.

The dipolar interaction can also be conveniently written using the irreducible spherical
tensors as 
\begin{equation}\label{ddham}
\hdd = \omega_{d}\sum\limits_{m=-2}^{2} (-1)^m Y_{-m}^2(\theta,\phi) T_{m}^2 =
\sum\limits_{m=-2}^{2} \omega_{d,m} T_{m}^2, 
\end{equation}
where,
$Y_{-m}^2(\theta,\phi)$ is the spherical harmonics of rank $2$ and order $-m$,
$T_{m}^2$ is the irreducible spherical tensor of rank $2$ and order $m$,
and $\omega_{d,m} = (-1)^m Y_{-m}^2(\theta,\phi) \omega_{d}$ is an abbreviated form introduced for notional
simplicity.
$\theta$ and $\phi$ are the polar and azimuthal angle of the orientation of the dipolar vector
\textit{w.r.t.} the direction of the polarization of the spins, respectively \cite{smith1992a, smith1992b}. 
We also have $\omega_{d} = \sqrt{\frac{6\pi}{5}}\frac{\gamma^2\hbar}{r^3}$, where $\gamma$ is the gyromagnetic
ratio of the spin and $r$ is the distance between the spins.

We must note that
the assumption of a complete dipolar interaction between the qubits, as given above, instead of a simplified
$\vec{\sigma}^{1}.\vec{\sigma}^{2}$ secular form results in certain differences in the dynamics of the qubits.
First of all, the secular form (contains only the $T_{0}^2$ term) is a part of the full form given above.
In the interaction representation, $m\neq 0$ terms are the time-dependent part of dipolar coupling. These
terms contribute to the dissipator and result 
in the reduction of the persistent entanglement of
the qubit pairs.  Other differences which results from $T_{m}^2\ \forall\ m\neq 0$ would be
discussed in detail in the section \ref{equations}. 



Finally, $\hl(t)$ denotes time-dependent thermal fluctuations acting on the environment. Since the thermal
fluctuations are ubiquitous in an environment, we explicitly take this into account along with the other
terms in the Hamiltonian. Since the thermal fluctuations do not destroy the equilibrium of the environment,
as a simple model, we assume that $\hl(t)$ commutes with the static Hamiltonian of the environment
$\Hq^{\circ}_L$. It is also assumed that the timescale of fluctuations is much faster than the timescale of
the system. The explicit form of the fluctuation Hamiltonian is given by
\begin{eqnarray}
\Lhl(t) = \sum_i f_i(t)\vert \phi_i \rangle \langle \phi_i \vert
\end{eqnarray}
where $\{\vert\phi_i\rangle\}$ are the eigenstates of $\Hq_{L}^{\circ},$ $i$ is the number of energy levels of the
environment.  Here,
$f_i$ is modeled as the stationary and delta-correlated, Gaussian stochastic variables with standard
deviation $K$. Therefore, $\overline{f_i(t)}=0$, $\overline{f_i(t_a)f_j(t_b)}= \frac{1}{\tau_c} \delta_{ij}
\delta(t_a-t_b)$ and $K^2/2 = 1/\tau_c$.

\section{Equations of motion}
\label{equations}
To arrive at the equation of motion from the above Hamiltonian, we use recently-introduced
fluctuation-regulated quantum master equation, whose form is given by,
\begin{eqnarray}\label{frqme}
\frac{d\rho_s}{dt}&=& -i\, \tr_{L}\Big[\heff(t),\rh(t)\otimes\rl\Big]^{sec}\nn\\
&&-\int\limits^{\infty}_0 d\tau\, \tr_{L}\Big[\heff(t),\Big[\heff(t-\tau),
\rh(t)\otimes\rl\Big]\Big]^{sec}e^{-\frac{\vert\tau\vert}{\tau_c}}
\label{eqa-1}
\end{eqnarray}
where, $\heff$ contains the system-environment coupling and other interactions. We note that the effect of
the fluctuations are contained in the kernel $e^{-\frac{\vert\tau\vert}{\tau_c}}$ which is obtained from a
cumulant expansion of the propagator containing the fluctuations. The equation is written in the interaction
representation \textit{w.r.t.} to the static Hamiltonians of the system and the environment. Since, FRQME is
relatively new, hence we provide a brief summary of its derivation in the appendix \ref{derive-frqme}.



We note that the spectral density functions of the system-environment coupling are obtained as
\begin{eqnarray}
\int\limits_0^{\infty} d\tau e^{-\tau/\tau_c} 
\langle B_{\rm int}(t)B_{\rm int}^{\star}(t-\tau)\rangle
\trl\{\mathcal{L}_{\pm}\mathcal{L}_{\mp}\rho_{L}^{eq}\}
\langle \eta(\vec{r}_n)\eta^{\star}(\vec{r}_m)\rangle
\end{eqnarray}
where, $B_{\rm int}(t)$ is the time-dependent scalar part of the system-environment coupling in the
interaction representation and we write the Fourier transform of the time correlation 
$\langle B_{\rm int}(t)B_{\rm int}^{\star}(t-\tau)\rangle = J \pm i\delta\omega$, with $J$ and
$\delta\omega$ denoting the real and the imaginary parts.
%
%
The term $\langle \eta(\vec{r}_n)\eta(\vec{r}_m)\rangle = \alpha^{(1-\delta_{nm})}$ indicates the spatial correlation of the bath,
We note that the scalar factor $\alpha \in
[0,1]$ appears only in the \emph{cross} terms, when the environment couples two different spins ($n\neq m$)
through the environment operator. 
For the operator part, we assume that $\trl\{\mathcal{L}_{\pm}\mathcal{L}{\mp}\rho_{L}^{eq}\} = 1 \pm M_{\circ}$
where $\rho_{L}^{eq}$ is the equilibrium density matrix of the environment. This assumption implies that the qubits will inherit
equilibrium populations $(1\pm M_{\circ})/2$, i.e., an equilibrium polarization
$\langle\frac{\sigma_z}{2}\rangle = M_{\circ}$ when the environment is completely local, i.e., $\alpha = 0.$
We note again that in the above, it is also assumed that only energy levels of the environment contribute to the above
integral whose separation exactly matches with the Larmor frequency of qubits i.e. the resonant modes. 

A common environment could be modeled as a tight-binding chain of oscillators to which the qubits are
coupled. For such cases, $\alpha$ had been shown to be a decreasing function of the distance between the
qubits $r$. It could be chosen to have an exponential form $\alpha = \exp(-r/\xi)$, where $\xi$ is the
correlation length of the spatially correlated environment originating from the chain of oscillators
\cite{jeske2013}. In such a formulation, $\alpha$ remains a measure of the \emph{commonness} of the
environment. We refer to the works of McCutcheon \etal for more details on the spatially correlated
environment \cite{mccutcheon2009}.

We note that $\alpha = 0$ signifies a separate local environment for each qubit. With $\alpha=0,$ the
qubits relax to their equilibrium polarization with a time constant of $1/J.$ The $\delta\omega$ terms in the
spectral density give rise to the Lamb shifts in equations of the observables that do not commute with
$\sigma_z^{(1,2)}$.  On the other hand, $\alpha = 1$ indicates that both the spins are coupled to a single common
environment. For intermediate values of $\alpha$, the environment is a mixture of the common and the
local environments. Interestingly, since $\alpha$ and the dipolar coupling both are functions of $r$, we
expect competition between the two distance-dependent processes and investigate that in this manuscript. 

Using FRQME, we obtain the dynamical equation as,
\begin{eqnarray}\label{lindblad}
\frac{d \rho_s}{dt}&=& -i\big[ \omega_{d,0} T^2_0+ H_{\rm Lamb}+H_{\rm dds}, \rho_s\big] 
+ \mathcal{D} \rho_s + \mathcal{Q} \rho_s
\end{eqnarray}
where, $H_{\rm Lamb}$ is the Lamb shift from the system-environment coupling, and is given by,
\begin{eqnarray}
H_{\rm Lamb}&=& -\sum\limits_{i,j=1}^2 \alpha^{(1-\delta_{ij})}\delta\omega\left( (1+M_{\circ}) \sigma_{+}^{i}\sigma^{j}_{-} -
(1-M_{\circ})\sigma_{-}^{i}\sigma_{+}^{j} \right),
\end{eqnarray}
where, $\alpha$ appears only in the cross terms ($i\neq j$).
$H_{\rm dds}$ is the second order shift similar to the Lamb shift term shown above, arises from the
$T^2_{m}, {m\neq 0}$ part of the dipolar coupling Hamiltonian, and is given by,
\begin{eqnarray}
H_{\rm dds} = - \sum\limits_{\substack{m=-2 \\ m\neq 0}}^{2} \delta\kappa_m T^2_{m}T^2_{-m}  
\end{eqnarray}
where, $\delta\kappa_m,$ $m \in \{1,2\}$ is the imaginary part of the spectral density corresponding to
$T^2_{m}$ term from $\hdd$. The complex spectral density corresponding to $T^2_{m}$ term from $\hdd$
is given by,
\begin{eqnarray}
\kappa_m + i\delta\kappa_m=
\frac{\vert\omega_{d,m}\vert^2 \tau_c}{1+(m\omega_{\circ}\tau_c)^2}\left(1 +
i m\omega_{\circ}\tau_c\right).
\end{eqnarray} 


The form of the dissipator $\mathcal{D}\rho_s$ from the system-environment coupling is obtained in the Lindbladian form as,
\begin{eqnarray}\label{DHsl}
\mathcal{D}\rho_s = \sum\limits_{i,j=1}^2 \alpha^{(1-\delta_{ij})} J \Big[ && (1+M_{\circ})\big(2 \sigma_-^i \rho_s \sigma_+^j 
-\{ \sigma_+^j \sigma_-^i , \rho_s\} \big) \nonumber \\
+ &&  (1-M_{\circ})\big(2 \sigma_+^i \rho_s \sigma_-^j 
-\{ \sigma_-^j \sigma_+^i , \rho_s\} \big)\Big]
\end{eqnarray}
The higher-order contributions of the dipolar Hamiltonian, denoted by 
$\mathcal{Q} \rho_s$ in the Eq. (\ref{lindblad}), is obtained in a Lindbladian form given by,
\begin{eqnarray}\label{DHdd}
\mathcal{Q} \rho_s &=& \sum_{m=-2}^{2}\,
\kappa_m\,\left[2T^2_{-m}\rh T^2_{m}
- \left\{T^2_{m}T^2_{-m},\rh \right\}\right].
\end{eqnarray} 
In case of dipolar interaction, the first order term consists of $m=0$ term (shown above in Eq.
\ref{lindblad}) and the second order contributions 
are coming from the following combination of $m$ value of $\hdd$. They are, $m = \{0,0\},\{-1,1\},\{-2,2\}$.
Other combinations do not survived under secular approximation \cite{cohen2004}. 




\subsection{Analysis in terms of observables}
The natural choice for analyzing the master equation described by Eq. (\ref{lindblad}) is to move to a Liouville
space description.  Liouville space presentation of this QME is, $\left[\frac{d \hat{\rho_s}}{dt} =
\mathcal{\hat{L}}\hat{\rho_s} \right]$ where, $\mathcal{\hat{L}}$ is the Liouvillian superoperator.  The
resulting Liouvillian matrix is a $n^2 \times n^2$ matrix, and the density matrix is a $n^2 \times 1$ column
matrix, where $n$ is the length of the Hilbert space. The closed-form expression of the Liouvillian is large
and hence is not convenient
for algebraic manipulation. As such, we recast the equations of motion in terms of the expectation values of
observables which are much more convenient.

We note that the Liouvillian is completely symmetric under an
exchange of the qubit indices 1 and 2. Instead of using 15 independent elements of the reduced density
matrix $\rho_s$, we use the observables' expectation values constructed from the Pauli matrices of the qubits.
Eq. (\ref{observables}) describe the construction of the observables and their expectation values.
We note that the positive (negative) signs generate a set of symmetric (asymmetric) observables with respect
to the exchange of qubit indices. It is clear that we have nine symmetric and six asymmetric observables,
which are given by,
\begin{eqnarray}
M_{\alpha}^{(\pm)} &=& \frac{1}{2} {\rm Tr}_s  [ (\sigma_{\alpha}\otimes \mathbb{I} \pm \mathbb{I} \otimes \sigma_{\alpha})\rho_s ] \nn\\
M_{\alpha\beta}^{(\pm)}&=& \frac{1}{4} {\rm Tr}_s [(\sigma_{\alpha} \otimes \sigma_{\beta} \pm \sigma_{\beta} \otimes \sigma_{\alpha})
\rho_s ],\quad \forall\ \alpha\neq\beta \nn \\
M_{\alpha\alpha} &=& \frac{1}{4} {\rm Tr}_s [(\sigma_{\alpha}\otimes \sigma_{\alpha})\rho_s] 
\label{observables}
\end{eqnarray}  
where, $\alpha,\beta \in \{x,y,z\}$. The symmetric observables are denoted without the superscript $(+)$
in the manuscript's remaining part, and the antisymmetric observables are denoted by $A$. 
Moreover, instead of using $M_{xx}$ and $M_{yy}$, we use $M_c = M_{xx} + M_{yy}$ and $A_c = M_{xx} -
M_{yy}$, for further simplifications.
In terms of observables, we can write the Eq. (\ref{lindblad}) as inhomogeneous first-order coupled linear 
differential equations. 
We construct the equations for all the above observables using Eq.
(\ref{lindblad}) and are shown in the appendix \ref{alleqs}. 

Of the fifteen observables, $M_z$ evolves to an equilibrium value due to the system-environment coupling.
Naturally, all observables which are coupled to $M_z$ would also acquire finite equilibrium value. As such,
we investigate only the set of coupled equations containing $M_z$. The other symmetric and asymmetric
observables have zero value in the steady-state and are not important for the storage problem.
The coupled equations of interest are given by, 
\begin{equation}\label{3eq}
\frac{1}{J}\left[\begin{array}{c} \dot{M}_z \\ \dot{M}_{zz}\\ \dot{M}_{c} \end{array} \right] = 
\begin{bmatrix}
-(2 + \kappa_1^{\star} + 4 \kappa_2^{\star}) & 0 & 4M_{\circ}\alpha \\   M_{\circ} & -(4 + 2
\kappa_1^{\star}) & 2 \alpha + \kappa_1^{\star}\\ 
-M_{\circ} \alpha  & 4 \alpha + 2 \kappa_1^{\star} & -(2 + \kappa_1^{\star}) \end{bmatrix}  \left[\begin{array}{c} M_z  \\
M_{zz}\\ M_c \end{array}\right] + \left[\begin{array}{c} 2M_{\circ} \\  0\\ 0 \end{array}\right]
\end{equation} 
where, $\kappa_m^{\star}=\kappa_m/J$. We choose to represent the time in the units of $1/J$, such that
the effect of the dipolar coupling could be investigated using the scaled variables $\kappa_m^{\star}$.
In the dynamical equation of the observables $\{M_z, M_{zz}, M_c\}$, there are no first-order terms corresponding to $\hdd$.
We note that only the non-secular ($m \neq 0$) higher-order terms from the dipolar coupling ($\kappa_1$ and $\kappa_2$) 
feature in the above equations. 
Interestingly, the term
$\kappa_1$, which originates from $m=1$ terms of the dipolar Hamiltonian, couples the observables $M_{zz}$ and $M_c$.


\comment{Note:\\
density matrix could be written in terms of these observables. Show.

Note:\\
the singlet is Mzz + Mc, and triplet is Mzz - Mc. The former does not evolve for alpha = 1. See notes. 
}

\section{Results}
\label{results}

The square matrix in Eq. (\ref{3eq}) is non-singular for $\alpha \neq 1$. In this case, the solution does not have any initial value 
dependence. The steady-state solution is given by,
\begin{eqnarray}\label{3sol-neq1}
M_z\sts &=& \frac{2 M_{\circ}(1+\alpha + \kappa_1^{\star})}{C_1}\nn\\
M_c\sts &=& \frac{M_{\circ}^2 \kappa_1^{\star}}{2C_1}\nn\\
M_{zz}\sts &=& \frac{M_{\circ}^2(2+2\alpha + \kappa_1^{\star})}{4C_1}
\end{eqnarray} 
where, $C_1 = (1+\kappa_1^{\star})(2+\kappa_1^{\star}+4\kappa_2^{\star})+\alpha(2+\kappa_1^{\star}+4\kappa_2^{\star} -
\kappa_1^{\star}M_{\circ}^2)$.  In the absence of the 
dipolar-coupling, i.e. $\kappa_1^{\star}=\kappa_2^{\star}=0$, the steady state solution reduces to, $M_z\sts = M_{\circ}$, $M_c\sts = 0$ 
and $M_{zz} \sts =M_{\circ}^2/4$. 
This result matches exactly with the steady-state value obtained for the uncoupled qubits interacting with the regular
environment \cite{benatti2006b, mccutcheon2009}.

On the other hand, for $\alpha = 1$, there exist a superoperator $\hat{D} = D\otimes
\mathbb{I}-\mathbb{I}\otimes D^{T}$, where, $D = \sigma_{x}\otimes \sigma_{x} + \sigma_{y} \otimes
\sigma_{y} + \sigma_{z} \otimes \sigma_{z}$. We note that $\hat{D}$ commutes with the dissipators in Eq. (\ref{lindblad})
Naturally, it implies that there exist a conserved quantity,
\begin{eqnarray}\label{conserved}
\frac{d}{dt}\left( M_{xx}+M_{yy}+M_{zz} \right) = 0.
\end{eqnarray}   
Therefore, the final solution has an initial value dependence. The steady-state
dynamics is confined between the observables-$\{M_z,M_{zz},M_c\}$. In the Eq.~(\ref{3eq}), the matrix is singular
for \mbox{$\alpha=1$}. The singularity also implies that the final steady state should have an initial value
dependency.  The steady state solution is given by,
\begin{eqnarray}\label{3sol-eq1}
M_z\sts &=& \frac{2M_{\circ}(3+4F)(2+\kappa_1^{\star})}{C_2}\nn\\
M_c \sts &=& \frac{-2 M_{\circ}^2+2F(2+\kappa_1^{\star})(2+\kappa_1^{\star}+4\kappa_2^{\star})}{C_2}\nn\\
M_{zz} \sts &=& F - M_c \sts
\end{eqnarray}
Here, $C_2 = 4M_{\circ}^2 + 3(2 + \kappa_1^{\star})(2 + \kappa_1^{\star} + 4\kappa_2^{\star})$, and $F =
(M_{xx}+M_{yy}+M_{zz})\stoo$. In the absence of the dipolar coupling, i.e. for $\kappa_1^{\star}=k_2^{\star}=0$, the
above equations reduce to $M_z\sts = M_{\circ}(3+4F)/(3+M_{\circ}^2)$ and $M_c\sts =
(4F-M_{\circ}^2)/(6+2M_{\circ}^2)$, $M_{zz} \sts = F - M_c\sts$. This result is in agreement with earlier work \cite{benatti2006b,mccutcheon2009}.

\begin{figure*}[htb]
\raisebox{3cm}{\normalsize{\textbf{(a)}}}\hspace*{-1mm}
\includegraphics[width=0.40\linewidth]{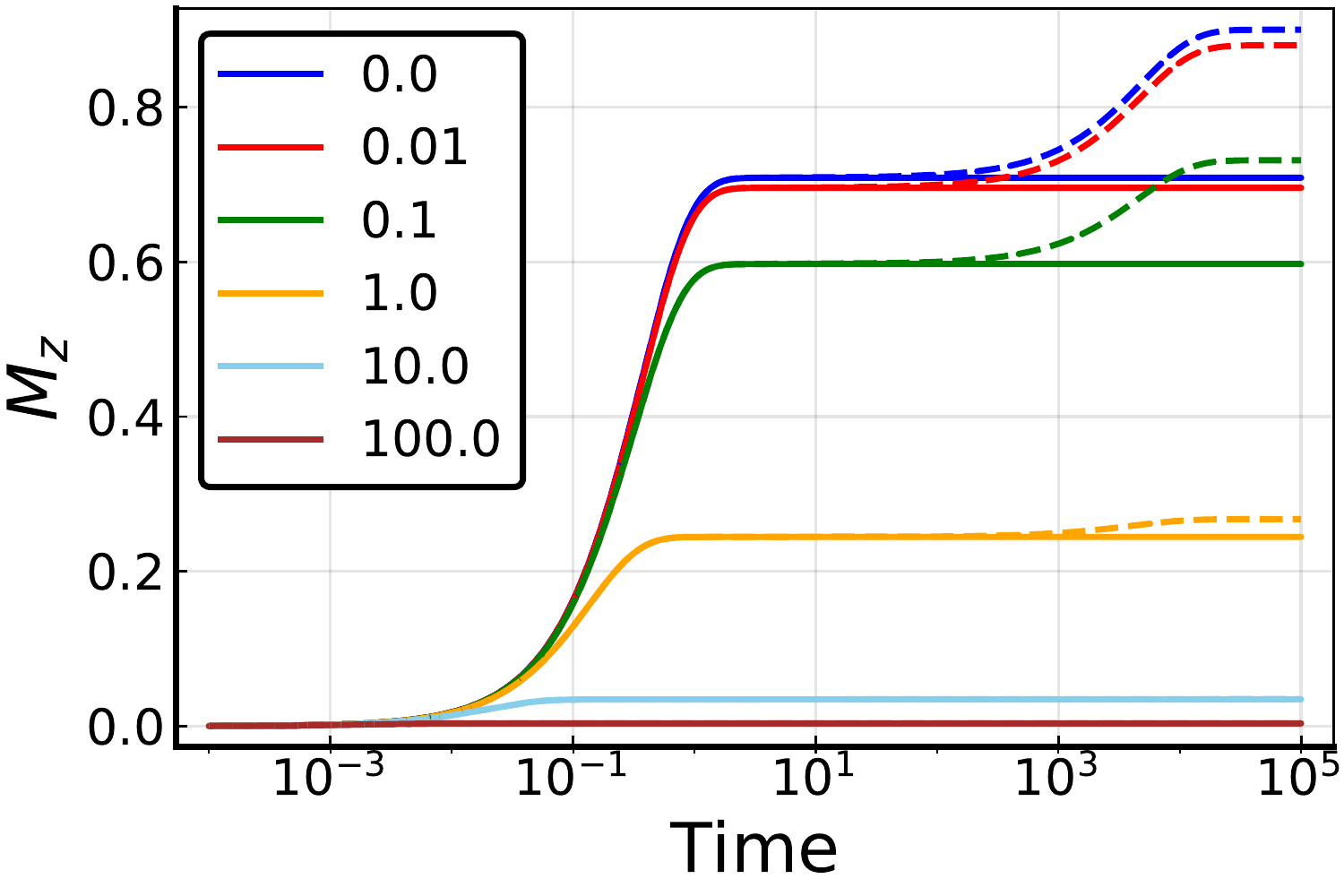} 
\hspace*{4mm}
\raisebox{3cm}{\normalsize{\textbf{(b)}}}\hspace*{-1mm}
\includegraphics[width=0.40\linewidth]{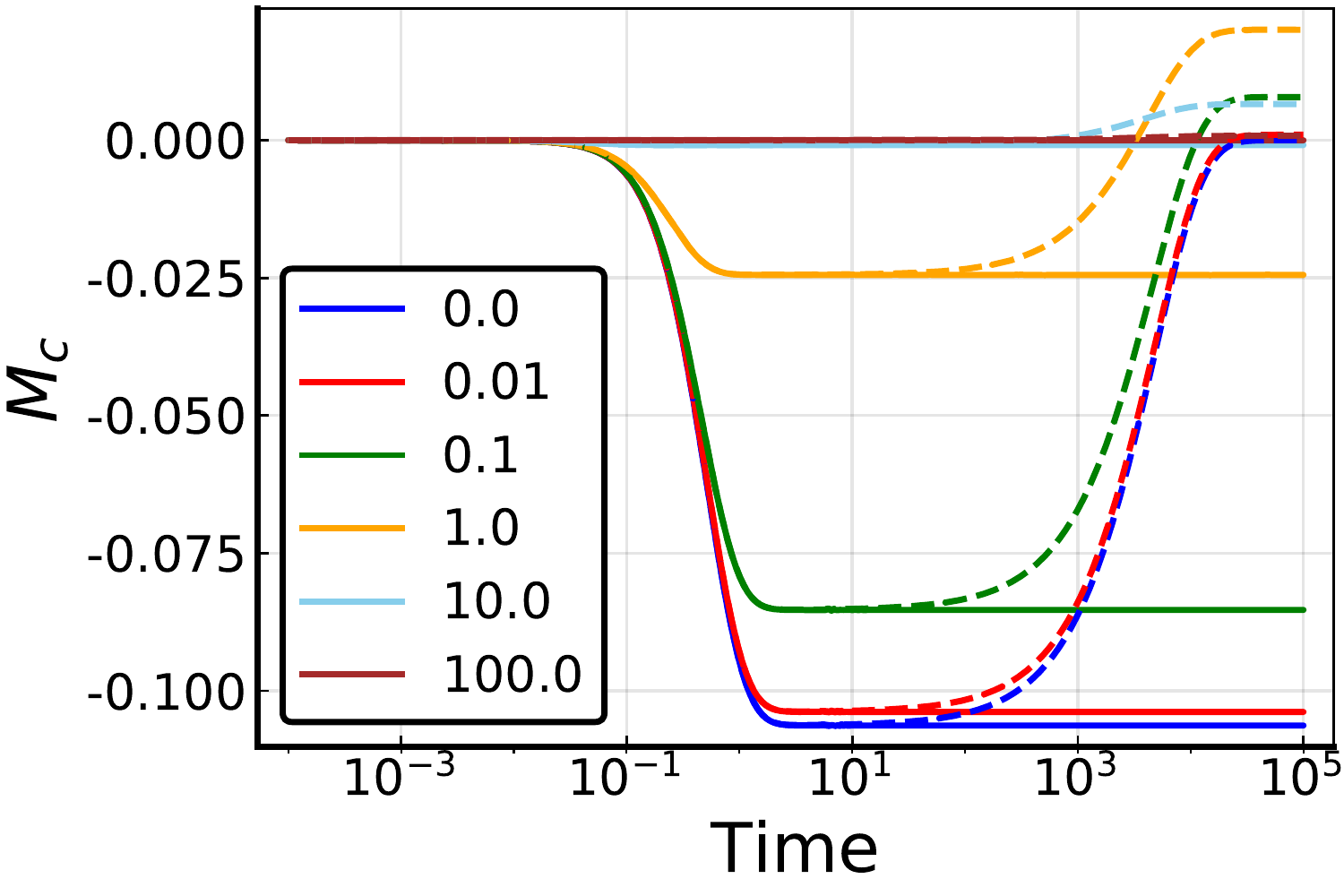}
\hspace*{4mm}
\raisebox{3cm}{\normalsize{\textbf{(c)}}}\hspace*{-1mm}
\includegraphics[width=0.42\linewidth]{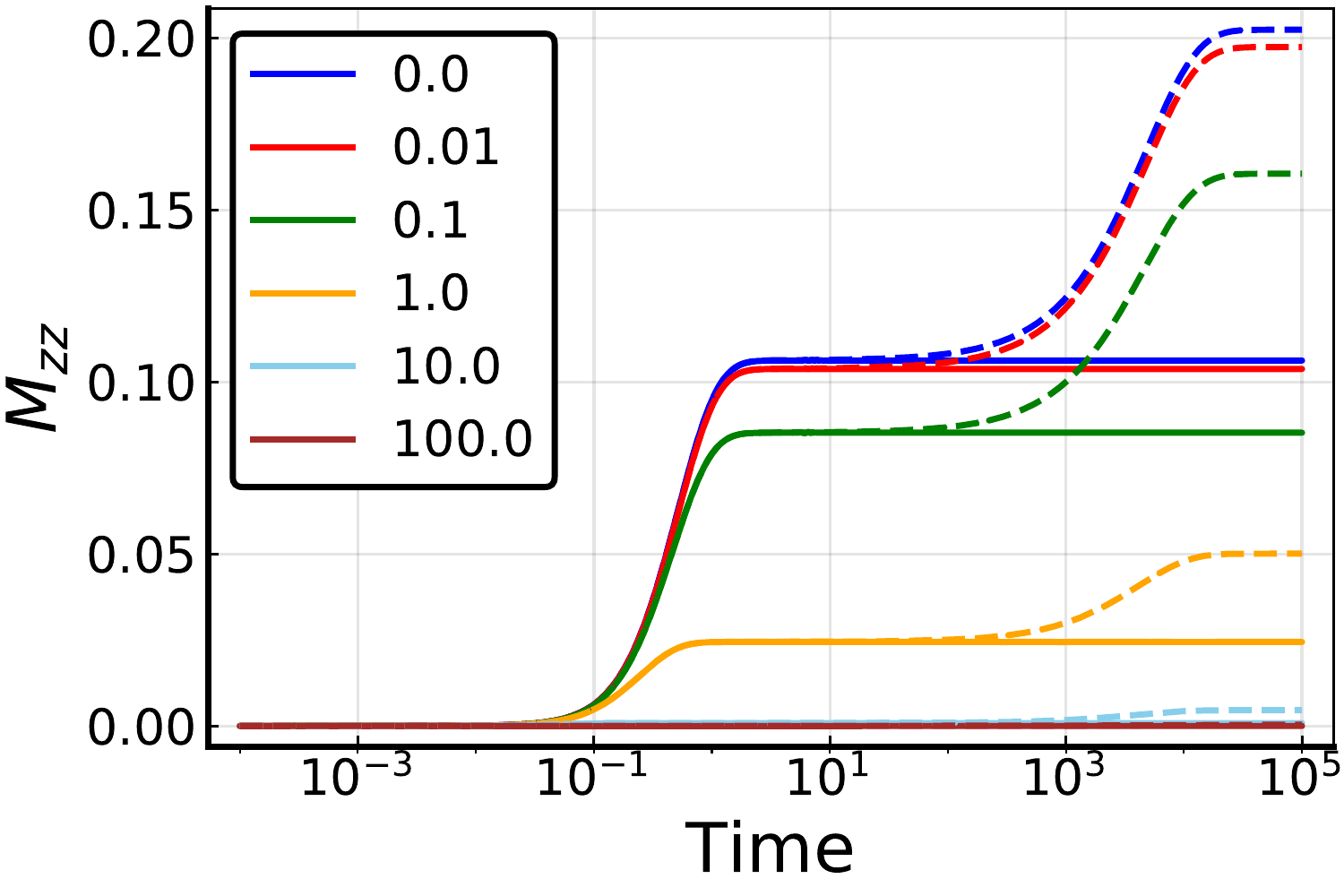} 
\caption{(a), (b) and (c) show the behaviors of three observables $M_z$,  $M_c$, and $M_{zz}$, respectively, as a function of
time in the units of $1/J$. To arrive at these plots, the equation (\ref{3eq})
has been solved numerically. For each variable, the relative
strength of the dipolar interactions and the system-environment coupling was varied by adjusting the ratio
$\kappa_1^{\star} = \{0, 0.01, 0.1, 1.0, 10, 100\}.$ The solid lines show the solutions for $\alpha = 1$ and the dashed
lines show the solutions for $\alpha = 0.9999$. In the presence of a completely common environment $\alpha =
1$, the system reaches a steady state in agreement with the solutions in Eq. (\ref{3sol-eq1}). For $\alpha\neq 1$, 
the solutions are depicted by dashed lines and the steady states are given by Eq. (\ref{3sol-neq1}). We also
note the presence of a slow timescale in addition to the fast timescale present for $\alpha=1$ case.   
The fixed parameter sets are chosen as $\{M_{\circ} = 0.9, \kappa_2^{\star}=\kappa_1^{\star}\}$. 
} 
\label{fig-1}
\end{figure*}

Although Eq (\ref{3eq}) appears simple, its general solution as a function of $t$ in a closed-form is
algebraically cumbersome. So, we show the numerical solutions of these equations in graphical form.  Figures
\ref{fig-1}(a-c) depict the behaviors of $M_{z}$, $M_c$, and $M_{zz}$ as a function of time (in the scaled
units $t \to J t$) for various values of $\kappa_1^{\star}$ as shown in the legends. For these plots, it was
assumed that $\kappa_2 = \kappa_1$.  The solid lines show the behavior for $\alpha = 1$. We note that since
$M_{zz} + M_c$ is a constant of motion for $\alpha = 1$, each of these observables reaches a steady state.
The steady-state values are in agreement with the analytical forms given above and show a marked drop in the
value of $M_c$ as we increase the dipolar strength. For $\alpha \neq 1$, we obtain different dynamics with
the presence of a slower timescale with which the system reaches an eventual steady state. From Eq.
(\ref{3eq}), we note that $\dot{M}_{zz}+\dot{M}_c \propto (1-\alpha)$ and hence the slower timescale depends
on the departure from the commonness of the environment. This critical slowing down of the dynamics has
been reported before, although its dependence on the dipolar strength remains previously unexplored
\cite{mccutcheon2009}.

From an experimental point of view, $\alpha$ tending to the value $1$ would be a much more
realistic scenario than the well-explored $\alpha = 1$ behavior. We observed that $M_c$ reaches the absolute
maximum at the scaled time $\sim 1$ (corresponding to an actual time $1/J$), and the value remains for one or two
orders of magnitude until it reaches its final steady-state value with a timescale proportional to
$(1-\alpha)$. Now $M_c$ is the expectation value of $\frac{1}{4}(\sigma_{x}\otimes \sigma_{x} + \sigma_{y} \otimes
\sigma_{y})$ of which singlet state is the eigenstate having the lowest eigenvalue. This leads to the straightforward result that
the persistence of $M_c$ provides us with a long-lived entanglement in the form of a singlet state. In the
next section, we explore the dynamics of this entanglement in more detail.



\subsection{Environment-induced entanglement}
\label{EIE}
Entanglement is entirely a quantum mechanical phenomenon of a multipartite system. If a single qubit only
interact with it's local environment, the final steady state is a thermal state, and a thermal state is a
non-entangled state. On the other-hand the EIE originating from a CE is a well-known phenomenon since the
seminal work of Plenio and others \cite{plenio1999, an_entanglement_2007, benatti_environment_2008,
zhang_entanglement_2007, choi2007}. We study the growth and the preservation of entanglements in the system
through a well-established measure concurrence
\cite{wootters1998,horodecki2009}. For a bipartite system, the concurrence as a measure was first proposed by
Wooters using the eigenvalues of the following Hermitian matrix $\sqrt{\sqrt{\rho_s} \tilde{\rho_s}
\sqrt{\rho_s}}$,  where $\rho_s$ is the given system density matrix and $\tilde{\rho_s}=
\sigma_y\otimes\sigma_y\, \rho^{\star}\,\sigma_y \otimes \sigma_y$. Alternatively, the eigenvalues can be
calculated from the non-Hermitian matrix $\rho_s \tilde{\rho_s} $. If $\lambda_i$ are the
eigenvalues arranged in the decreasing order, then the concurrence $(C(\rho_s ))$ is defined as,
\begin{eqnarray}
C( \rho_s ) &=& \text{max}\{0, \lambda_1-\lambda_2-\lambda_3-\lambda_4\}
\end{eqnarray} 

If $C( \rho_s )> 0$, then the system has entanglement. For $C( \rho_s )= 0$, we have a separable system.
In terms of the observables $\{M_z, M_{zz}, M_c\}$, the expression for the steady state entanglement is given by,
\begin{eqnarray}\label{concurrence}
C( \rho_s ) = \text{max}\{0 ,  2\vert M_c\vert - \frac12\sqrt{(1 + 4 M_{zz})^2 - 4 M_z^2}\}.
\end{eqnarray}
The above Eq. (\ref{concurrence}) corroborates our earlier assertion that there is a possibility of a persistent entanglement 
if there is a persistent zero quantum coherence, i.e., $M_c \neq 0$. But the converse is not true.

\begin{figure*}[h!]
\raisebox{3cm}{\normalsize{\textbf{(a)}}}\hspace*{-1mm}
\includegraphics[width=0.42\linewidth]{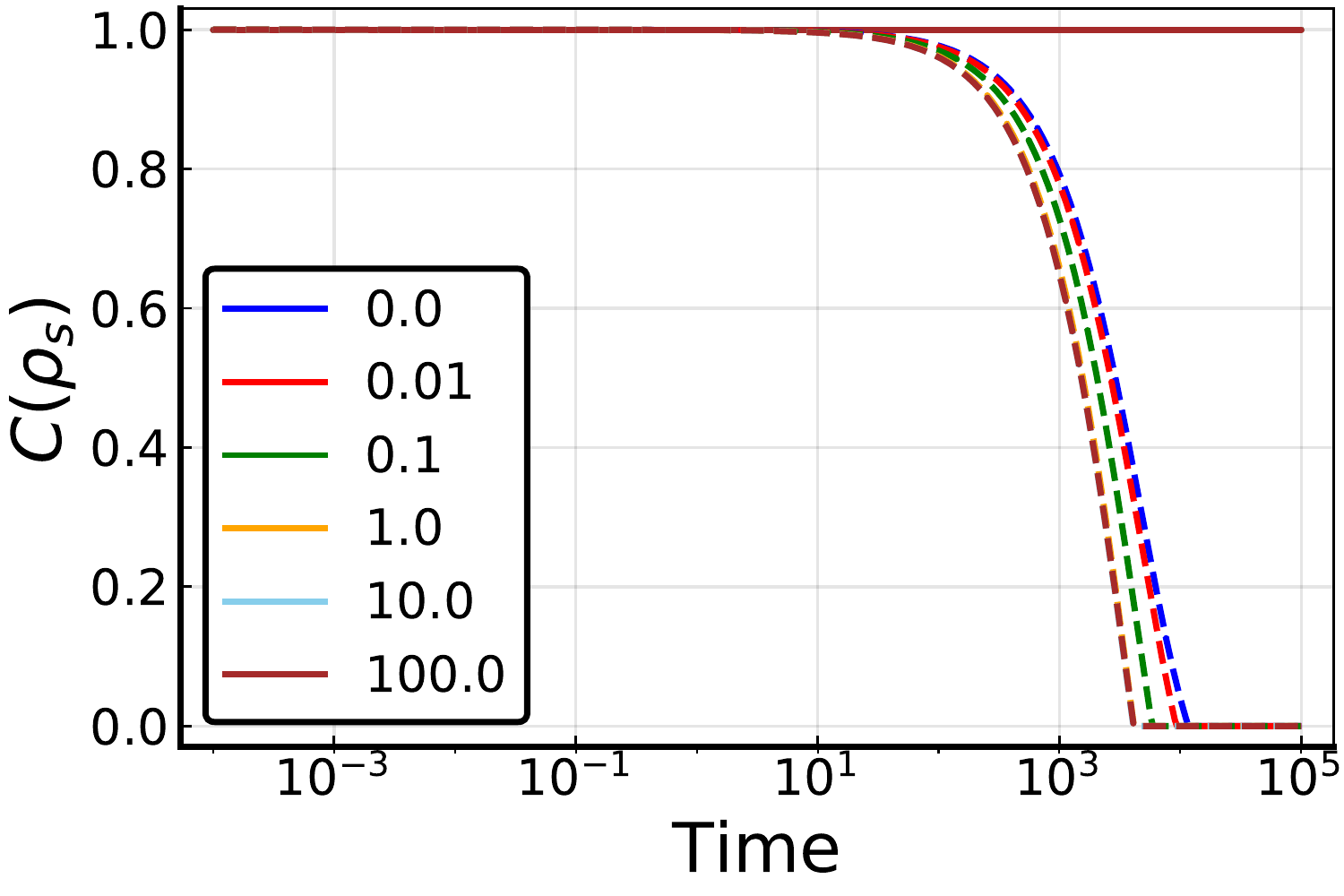} 
\hspace*{4mm}
\raisebox{3cm}{\normalsize{\textbf{(b)}}}\hspace*{-1mm}
\includegraphics[width=0.40\linewidth]{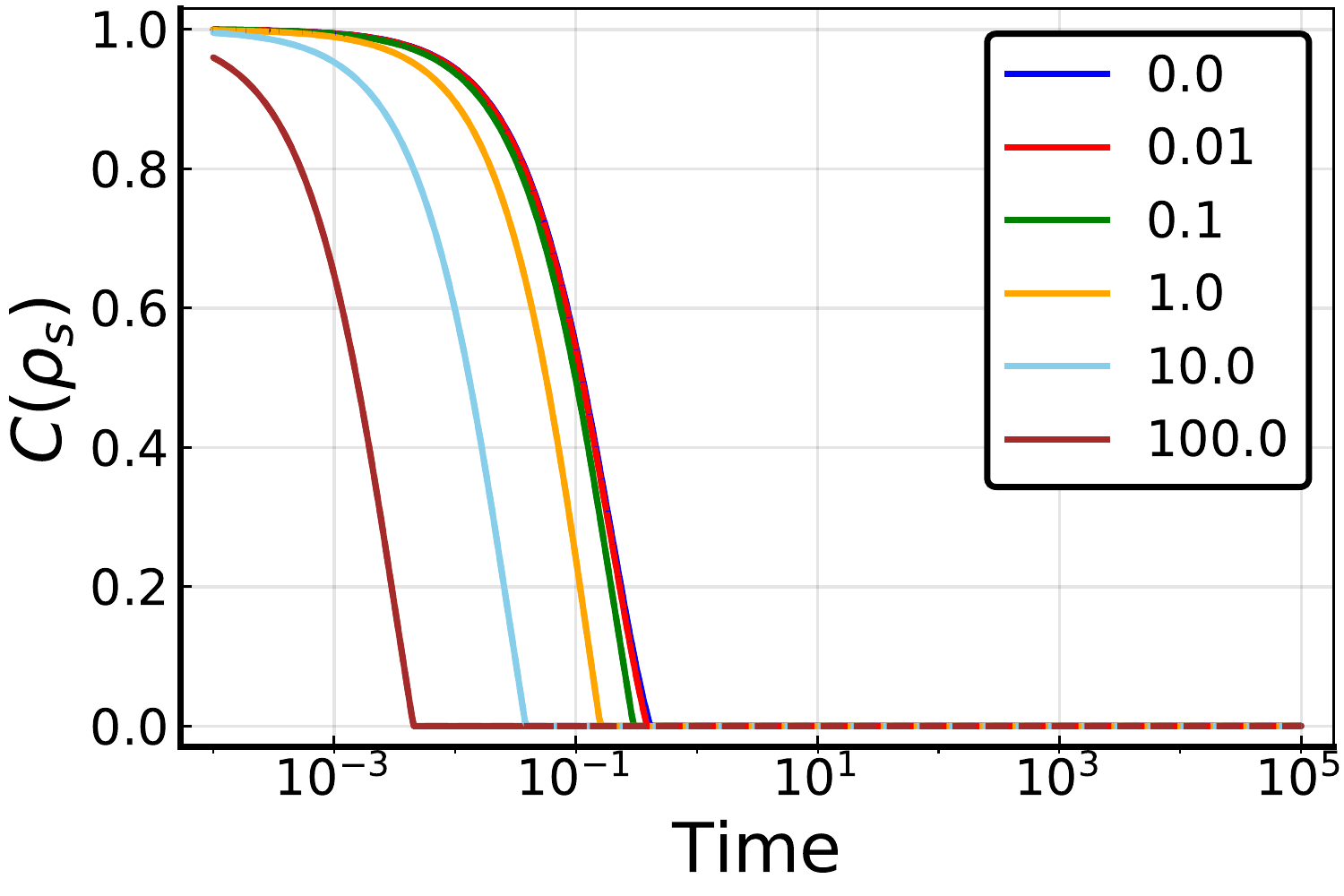}\\
\raisebox{3cm}{\normalsize{\textbf{(c)}}}\hspace*{-1mm}
\includegraphics[width=0.40\linewidth]{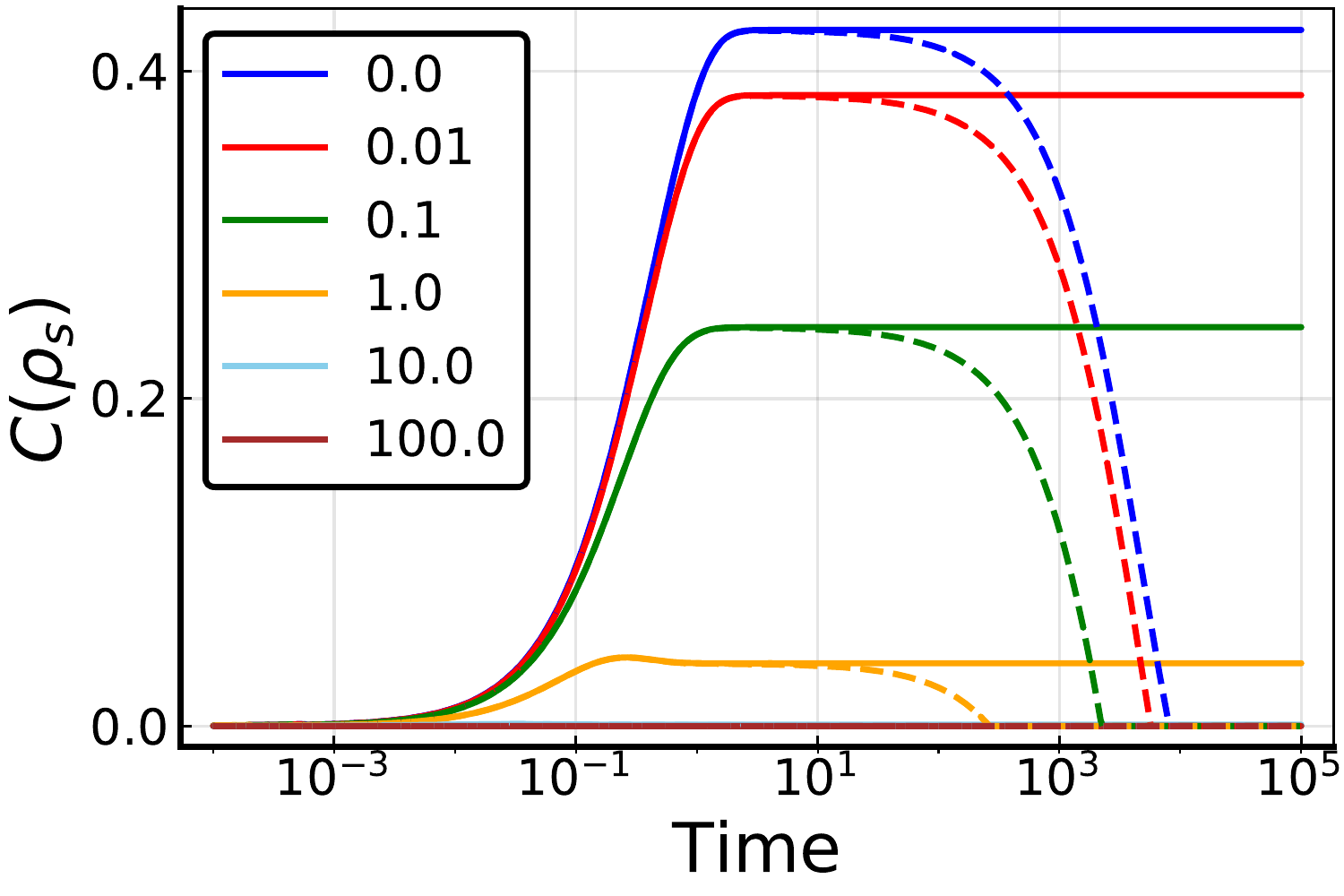}
\caption{Figure (a) shows the plot of concurrence $C(\rho_s)$ as a function of time for $\alpha=1$ by solid lines, 
and $\alpha = 0.9999$ by dashed lines, when the system was initially prepared in a singlet state. We observe
that the concurrence is perfectly preserved for $\alpha=1$ and decays to zero value for $\alpha < 1$. For
each $\alpha$ the dipolar coupling (in the scaled unit, $\kappa_m^{\star} = \kappa_m/J$) has been varied over 
four orders of magnitude from $1e-2$ to $1e+2$, as shown in the figure legends.   
Figure (b) shows the plot of concurrence $C(\rho_s)$ as a function of time for $\alpha=1$ by solid lines,
and $\alpha = 0.9999$ by dashed lines, when the system was initially prepared in a triplet state. 
We notice
that the triplet state decays almost four orders of magnitude faster than the singlet state. We also observe
that stronger dipolar coupling enhances the triplet decay rate.
Figure (c) shows the plot of concurrence $C(\rho_s)$ as a function of time for $\alpha=1$ by solid lines,
and $\alpha = 0.9999$ by dashed lines, when the system was initially prepared with zero values for all
observables. As $M_c$ builds up in the system, so does the concurrence according to Eq. (\ref{concurrence}).
However, the concurrence later decays for $\alpha<1$ (shown by dashed lines, as opposed to the solid lines
for $\alpha = 1$) with a slower timescale. We note that the stronger dipolar coupling enhances the decay
rate. The concurrence is preserved for over one order of magnitude of time ($\sim 1/J$) for small dipolar
coupling.
} 
\label{fig-2}
\end{figure*}

Using Eq. (\ref{concurrence}), we simulate three situations in which the qubit pair may be used as a storage
of entanglement. We prepare the system in an initial singlet state $\frac{1}{\sqrt{2}}(\ket{01}-\ket{10})$
and solve the equations of motion as in Eq. (\ref{3eq}) for $\alpha = 1$ and for $\alpha=0.9999$. Since the
singlet state constitutes a decoherence-free subspace for the entire Liouvillian for $\alpha = 1$, hence it
remains perfectly preserved as shown by the solid lines in Fig. \ref{fig-2}(a). On the other hand, for
$\alpha = 0.9999$, the singlet state remains preserved for a long time, but eventually decays (the dashed
lines in Fig. \ref{fig-2}(a)). On the other hand, when we prepare the system in a triplet state
$\frac{1}{\sqrt{2}}(\ket{01}+\ket{10})$, irrespective of the values of $\alpha$, the 
entanglement decays to zero value at a much shorter time and more so when the dipolar coupling is stronger ($\kappa_1^{\star} =
100$ case decays faster than $\kappa_1^{\star} = 0.01$ case). The plots for the triplet case are shown in
Fig. \ref{fig-2}(b).

Also, as yet another method of generation and storage of entanglement, we prepare the system having a
negative dipolar order $M_{zz} = -1/4$. In this condition, we have a strong entanglement build up and perfect
preservation for $\alpha = 1$ (solid lines in Fig. \ref{fig-2}(c)). On the other hand, for $\alpha =
0.9999$, the entanglement grows to the same value as that of $\alpha=1$ case and remain preserved for
several orders of magnitude of time in terms of 1/J. The entanglement eventually decays to zero value. We
note that the timescale of preservation is strongly dependent on the dipolar coupling strength. For
$\kappa_1^{\star} = 1$, the entanglement build-up is small and decays faster than for $\kappa_1^{\star} =
0.01$. As such, for such storage protocols, the dipolar coupling plays a detrimental role. 

\begin{figure*}[h!]
\includegraphics[width=0.40\linewidth]{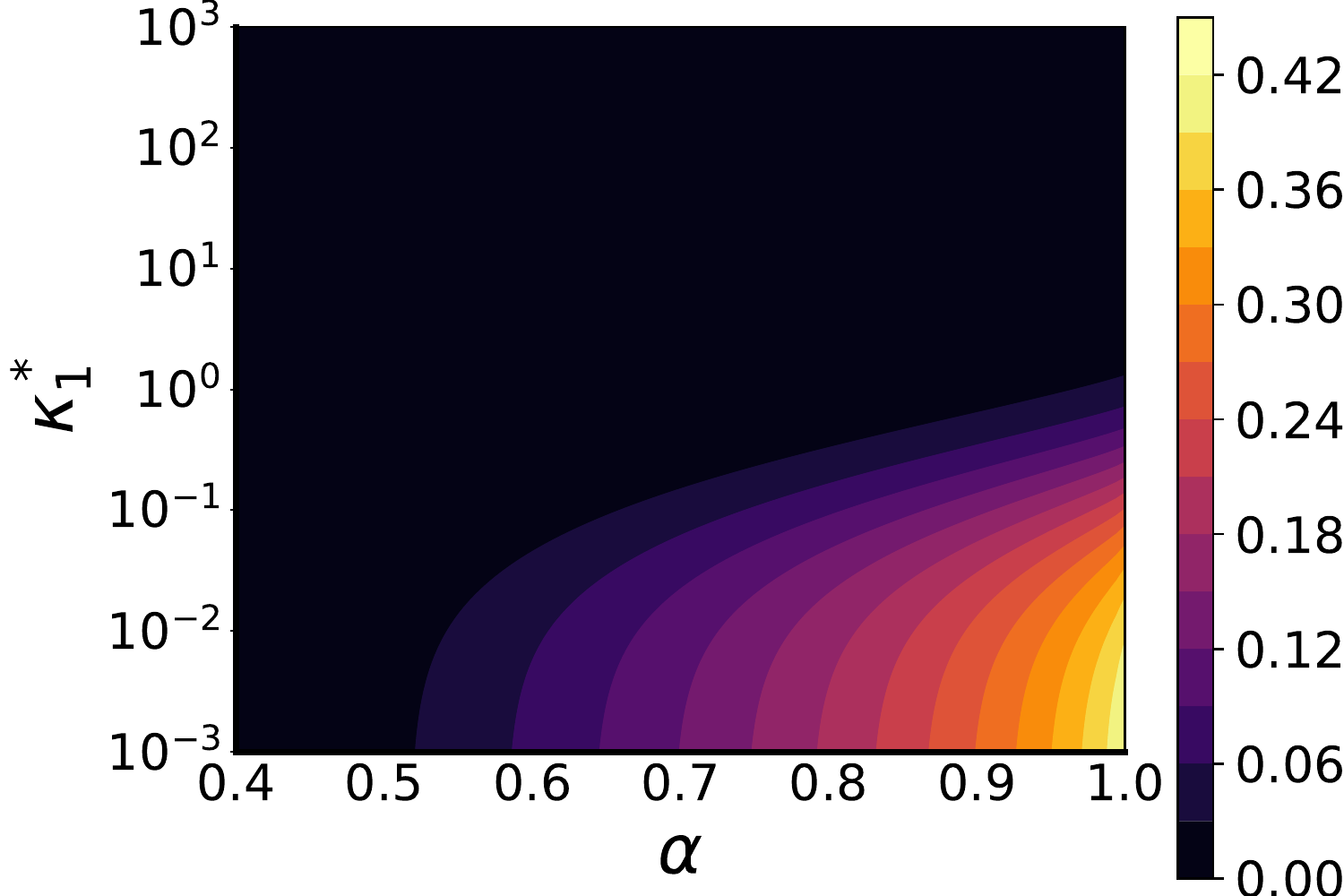}
\caption{The figure shows the filled contour of the maximum value of $C(\rho_s)$ a function of
$\kappa_1^{\star}$ and $\alpha$. The initial state has been chosen to be $M_z = M_c = 0$ and $M_{zz} =
-1/4$, which provides the maximum build-up of the entanglements due to a close-to-common environment with
$\alpha = 0.9999$.  To arrive at this plot, equations (\ref{3eq}) were solved for a specific choice of
$\{\kappa_1^{\star}, \alpha\}$ and the concurrence has been calculated as a function of time, and the
maximum value is selected. We note that the maximum value of the concurrence monotonically increases with increasing
$\alpha$, whereas, increasing the dipolar coupling amplitude, the entanglement between the qubits decays.
The color range that has been used in the plot is shown in the color bar on the right with the upper and lower
bounds of the range. We observe that with the increase of the dipolar coupling, the maximum value of the
concurrence decreases.
}
\label{fig-3}
\end{figure*}

To understand the roles of the commonness of the environment $\alpha$ and the effects of the nonsecular
dipolar coupling $\kappa_1^{\star}$, we plot the maximum value of the concurrence achieved by using the 
above protocol, for a range of values of $\{\alpha, \kappa_1^{\star} \}$. The resulting contour plot is
shown in Fig. \ref{fig-3}. The contours clearly show that the concurrence is highest for high values of 
$\alpha$ and low values of $\kappa_1^{\star}$. 

\comment{
A long ago, in 1953, Overhauser introduced the concept of nuclear polarization transfer using the dipolar
cross-relaxation effect \cite{overhauser1953_a,overhauser1953_b}. The experimental verification of the
Overhauser effect in case of metals is provided by the Slichter \etal \cite{slicter1953} and for non-metals
by Beljers \etal \cite{beljers1954}. The dipolar cross-relaxation process is actually determined by the
second-order contribution of the dipolar Hamiltonian. In the case of a dipolar coupled system in a dilute liquid,
the angular $\{\theta, \phi \}$ dependence of the dipolar interaction is random. The ensemble average of all
the spherical angle of the second-order terms gives an effective exponential decay. Solomon and Kubo \etal gave the rigorous
mathematical foundation of the cross-relaxation phenomena
\cite{solomon1955, kubo1954}. Since then, the dipolar relaxation process has been extensively used in the liquid state
Nuclear magnetic resonance (NMR) spectroscopy for $ T_1,\, T_2$ measurement \cite{tropp1980}. Recently
current authors analytically showed that for the solid-state system, using FRQME, one can expect a similar kind of
decay originating from local environment fluctuation \cite{saha2022}. Nevertheless, the decay rates for this case depend on the spherical angles. In this manuscript, our main motivation lies in the effect of such
second-order contribution of the dipolar interaction in the presence of the thermal fluctuation in the EIE. 
}

\section{Discussions}
\label{discussions}

Fig. \ref{fig-1}(b) shows that $M_c$ is finite for $\alpha = 1$.
So, the possibility of persistent EIE only occurs at the critical point $\alpha = 1$. For $0 < \alpha < 1$,
the EIE survives at the intermediate time-scale. Fig. \ref{fig-2}(c), captures the time-evolution of
$C(\rho_s)$ for $\alpha = 0.9999$ and the initial density matrix $\rho_s^{ini} = \mathds{1}/4$ $-
\left(\sigma_z \otimes \sigma_z\right)/4$. The EIE, as measured via concurrence, grows and attains maximum value in
the intermediate time and finally decays to zero at the steady state. The characteristics of the graph is same as
$\vert M_c \vert$ vs time in the Fig. \ref{fig-1}(b). We also show that the value of the $C(\rho_s)$
decreases for an increase in the dipolar interaction in Fig. \ref{fig-2}(c). The second-order dipolar terms
dominate over system-environment terms for larger $\kappa_1^{\star}$, which results in a faster decay of the
entanglement. 

\comment{
We note that EIE is a temperature-dependent phenomenon since the bath-correlation length is a function of
the temperature. For higher
temperatures, the bath correlation length becomes smaller, so it restricts the formation of EIE between the
separated qubits. In the lower temperature, the EIE becomes more prominent but in this temperature, the
system becomes highly non-Markovian \cite{Maniscalco2007}. Therefore in the Markovian regime, EIE can only
be generated by controlling the parameter, $r/\xi$ of the system. Zell \etal, McCutcheon \etal all reported
the distance sensitive behavior of EIE \cite{zell2009, mccutcheon2009}. For decreasing the inter-qubit
distances, the entanglement increases. It is expected that in this spatial limit ($r/\xi \to 0$), there may
exist such a $1/r^3$ dependent interaction between the qubits. We restrict ourselves in the perturbative
regime of the dipolar interaction. For strong inter-qubit coupling, the calculation is entirely different
from the perturbative one. Our results put a limitation to all the previously mentioned works on distance-sensitive EIE \cite{zell2009,mccutcheon2009}. 
}

In case of the dipolar-interaction, the secular term ($m=0$ in Eq. (\ref{ddham})) does not contribute in the time evolution of the following
observables $M_{xx}, M_{yy}, M_{zz}, M_z$. We note that the earlier versions of QME are based on the independent rate
approximation \cite{cohen2004}. In such formalism, the secular parts ($m\neq 0$) of the dipolar interactions
are taken into account, however secular part does not affect the evolution of the above observables. Since
the non-secular contributions had not been included, 
the previous approaches failed to capture the second-order effect of the
following pairs, $m = [\{0,0\}, \{1,-1\}, \{2,-2\} ]$ \cite{breuer2002}. In this context, FRQME is clearly a
better tool as it incorporates the effect of those terms in the dissipator. 
  
\comment{
The dipolar interaction between the qubits increases for decreasing $r$. The conserved quantity at
$r/\xi \to 0$ remains intact [$\dot{M_{xx}}+ \dot{M_{yy}} + \dot{M_{zz}}=0$] for any $\omega_d$. Therefore
we still get persistent EIE under the dipolar interaction. In terms of observables, the values of $M_z,
M_{zz}$ are reduced and $M_c$ becomes more positive. As a result, the concurrence $C(\rho_s)$ decreases. For
$\alpha<1$, McCutcheon \etal reported the lifetime of the entanglement is an inverse function of the
inter-atomic distance \cite{mccutcheon2009}. Here, our result suggests that (see Fig. \ref{fig-2}(c)) by
increasing the dipolar strength the lifetime of the entanglement decreases. So, the effective bath
correlation length decays due to the presence of the dipolar relaxation. Zell \etal shows that the EIE
occurs when the interatomic distance is less than a cutoff wavelength \cite{zell2009}. In the presence of weak
dipolar interaction, our work demonstrates that the cutoff wavelength is further reduced. 
}

The concepts for the generation and the preservation of EIE is extensively used in quantum storage devices,
in the conceptualization of quantum batteries, and environment mediated information transfer processes
\cite{Tabesh2020, Farina2019, kamin2020, gundogan2012}. Our results show that the dissipators from the
nonsecular part of dipolar coupling play an obstructive role in this process. It is clear from Fig.
\ref{fig-3} that the desirable condition for the maximum sustained concurrence is to reduce
$\kappa_1^{\star}$ and to increase $\alpha$.  One may increase the physical separation between the qubits to reduce $\kappa_1^{\star}$. At the same time, it is essential that qubits remain coupled to a common
environment with $\alpha \to 1$. Since $\alpha = \exp(-r/\xi)$, we must have $\xi \gg r$, i.e., an
environment with a large spatial correlation length much larger than the distance between the
qubits. This necessitates a suitably engineered quantum reservoir as a common environment. While there are
examples of reservoir engineering \cite{kurizki2015, myatt2000, rossini2007}, we note that Plenio \etal
originally reported coupling two physically distant atoms to a lossy cavity \cite{plenio1999}. Such methods
will be preferable to an environment created by a tight-binding chain of oscillators since a cavity is much
easier to fabricate.

We note that apparently $\kappa_1^{\star}$ could be reduced by increasing $J$ i.e., the system-environment
relaxation rate, since $\kappa_1^{\star} = \kappa_1/J$. This would indeed increase the concurrence, as evident in Fig. \ref{fig-3}. However, the length of the time over which the entanglement is preserved
increases with $1/J$. So, any attempt to increase $J$ will result in a shortening of the entanglement
storage time. As such, increasing the qubit separation would be the preferred way to decrease the dipolar
effects on the decay of the entanglement.

McCutcheon \etal showed that one conceptualization of a spatially-correlated environment was to use a
chain of coupled harmonic oscillators. For such an environment, the correlation length of the environment
remains a function of the temperature \cite{mccutcheon2009}. It is expected that with the decrease of
temperature, the correlation length will increase. This facilitates creating efficient storage since the
separation of the qubits could be increased. However, we also note that at lower temperatures, the validity
of a Markovian quantum master equation may become questionable. Particularly, the crucial assumption of
the separation of timescales of the fluctuations and the system dynamics may not hold at very low
temperatures since the fluctuations would have a longer correlation time. As such, lowering the temperature may
benefit the storage devices, however, the present analysis would probably not be valid at lower
temperatures.

\comment{
The principles of EIE has substantial use in making of quantum battery, quantum-storage device and
environment mediated information transfer processes \cite{Tabesh2020,Farina2019,kamin2020,gundogan2012}.
Second-order dissipative terms of the dipolar interaction are playing an obstructive role in this process.
Here, we identified that how to get rid of dipolar relaxation in EIE. In Fig. \ref{fig-3}, the contour plot
of $max\{C(\rho_s)\}$ is shown as a function of $\omega_d/g$ and $\alpha$. In the regime of $\omega_d/g<1$,
the effect of dipolar interaction in $C(\rho_s)$ is nearly negligible. Therefore, tuning of the system-bath
coupling can eliminate the effect of dipolar relaxation for a fixed $r$. Engineered quantum reservoirs are
used to control the coupling between the system and bath modes \cite{kurizki2015,myatt2000,rossini2007}. In
this case, an engineered system-bath coupling can be introduced for getting maximum $  C(\rho_s) $ by
increasing the value of $g_k$. The value of the maximum entanglement doesn't depend on the value of
system-bath coupling, it depends on $M_{\circ}$. But for larger $g_k$, the effect of $\kappa, \kappa_1$ can
be neglected in the Eq.-\ref{3eq} so, the condition will be similar to the case $\{\alpha=1, \omega_d =0\}$
and from the Fig. \ref{fig-3}, the concurrence will be maximum. It was known that to make an efficient
quantum storage device, the important condition is $r<<\xi$. We show that it is not a sufficient condition.
One must also satisfy $\omega_d < \vert g_k\vert $ along with it.   
}

\section{Conclusion}
\label{conclusion}

We demonstrate the detrimental effect of the dipolar interaction on the generation and the preservation of
entanglement in a qubit pair that is coupled to a spatially-correlated common environment. Using the
formalism of FRQME, we have shown that the second-order contribution of the nonsecular component of the
dipolar interaction strongly affects the EIE, which is a major disadvantage for the environment-mediated
information storage device. We discuss the efficacy of various engineered environments in mitigating this
problem. Our results show that qubits coupled to a spatially correlated lossy cavity would be a preferred
solution for entanglement storage.

\section{Acknowledgments}
The authors thank Arpan Chatterjee and Arnab Chakrabarti for insightful discussions and helpful
suggestions. SS acknowledges University Grants Commission for a research fellowship (Student ID: MAY2018- 528071).
 
\appendix
\section{Fluctuation-regulated quantum master equation} 
\label{derive-frqme}

Since the fluctuation-regulated quantum master equation (FRQME) is relatively new, we provide here a brief sketch of
its derivation. For more details, the reader may consult the original work of Chakrabarti \etal
\cite{chakrabarti2018b}. 

We consider a driven-dissipative quantum system in the presence of thermal fluctuations. Since the thermal
fluctuations are ubiquitous, we explicitly introduce the fluctuations as a separate Hamiltonian. 
The total Hamiltonian of the system and the environment is written as,
\begin{eqnarray}
\label{re1}
\Hq =\Hq_{s}^{\circ} + \Hq_{L}^{\circ} +\Lhsl + \Lhloc + \Lhl(t).
\end{eqnarray}
In the above eq-\ref{re1} the first two terms represent the free Hamiltonian of the system and environment.
$\Lhsl$ is the system-environment coupling Hamiltonian. $\Lhloc$ is the local interaction present
in the system, e.g. external drive, dipolar interactions, etc. $\Lhl(t)$ is the thermal fluctuation
Hamiltonian. The form of $\Lhl(t)$ is given by,
\begin{eqnarray}
\Lhl(t) = \sum_i f_i(t)\vert \phi_i \rangle \langle \phi_i \vert
\end{eqnarray}   
where, $\phi$s are the eigenstates of $\Hq_{L}^{\circ},$ $i$ is the number of energy levels in the bath. Here,
$f_i$ is modeled as the stationary and delta-correlated, Gaussian stochastic variables with standard
deviation $K$. Therefore, $\overline{f_i(t)}=0$, $\overline{f_i(t_a)f_j(t_b)}= \frac{1}{\tau_c} \delta_{ij}
\delta(t_a-t_b)$ and $K^2/2 = 1/\tau_c$.  Hence, all the coherences in the local environment decays within a
time-scale $\tau_c$. For a Markovian system, the time-scale of the evolution should be much larger than
$\tau_c$. In this formalism, we use the coarse-grained prescription provided by Cohen-Tannoudji \etal to
construct a smooth dynamical equation for the system \cite{cohen2004}. The coarse-grained time-scale is
defined as $\Delta t$, which is assumed to be $\tau_c << \Delta t << T_1,T_2$. Here $T_1, T_2$ is the system
relaxation time-scale. Starting from the von-Neumann Liouville equation of the total density matrix
$\rho(t)$ in the interaction frame, the solution at the time-interval $t+\Delta t$ is given by,
\begin{eqnarray}
\label{re2}
\rho(t+\Delta t)  &=& \rho(t) -i\int\limits^{t+\Delta t}_t [\heff(t_1) + \hl(t_1), \rho(t_1)] \,dt_1 
\end{eqnarray}
Here, $\heff(t)$ is the interaction representation of $\Lhsl+\Lhloc$.  The dynamical equation for the system
density matrices $\rho_s$ can be constructed by taking the partial trace over bath variables on the both
side of the eq-\ref{re2}. The commutator including $\hl(t)$ will vanish by taking the partial trace. In the
right-hand side the density matrices can be written as, $\rho(t_1)=U(t_1,t)\rho(t)U^{\dagger}(t_1,t) $,
where, $U(t_1,t)$ is the time evolution operator. The following condition, $\tau_c << 1/\omega_{sl},
1/\omega_{loc}$ ensures that the time-scale of the fluctuations is much faster than the system-time
evolution. Hence, the time-evolution operator must be linear in $\heff$ but it contains all possible higher
order terms in $\hl(t)$. The expression of the finite propagator is given by \cite{chakrabarti2018b},
\begin{eqnarray}\label{finalU}
U(t_1,t) &=&  \ul(t_1,t)-i\int^{t_1}_t dt_2 \heff(t_2) \ul(t_2,t) 
\end{eqnarray}
Here, $\ul(t_1,t)= \mathcal{T}\exp(-i \int\limits_t^{t_1} \hl(t_2) \,dt_2)$. At the initial time $t$, the
system and bath is assumed to be uncorrelated., $\rho(t) = \rl\otimes \rho_s(t)$. This is known as
Born-approximation \cite{cohen2004}. By using the form of $U(t_1,t)$ in eq-\ref{re2} and taking ensemble
average over all $f_i$s, the exponential kernel arises in the second order terms of $\heff(t)$ 
$[\overline{U_L(t_1) \rho(t) U^{\dagger}_L(t_2)}=\rl\otimes \rho_s(t) e^{-\frac{t_1-t_2}{\tau_c}}]$. The final form of
coarse-grained, time-local Markovian master equation given by,
\begin{eqnarray}
\frac{d\rho_s}{dt}&=& -i\, \tr_{L}\Big[\heff(t),\rh(t)\otimes\rl\Big]^{sec}\nn\\
&&-\int\limits^{\infty}_0 d\tau\, \tr_{L}\Big[\heff(t),\Big[\heff(t-\tau),
\rh(t)\otimes\rl\Big]\Big]^{sec}e^{-\frac{\vert\tau\vert}{\tau_c}}
\label{re4}
\end{eqnarray}
The superscript `$sec$' denotes the secular approximation, which requires neglecting the contribution of the
higher oscillating term of the FRQME. The above form of FRQME can be reduced to the famous
Gorini-Kossakowski-Lindblad-Sudarsan (GKLS) form. Hence trace preservation, and complete positivity also
holds for FRQME \cite{chakrabarti2018b}. The main feature of the above eq-\ref{re4} is the presence of
an exponential kernel in the second-order terms, which gives a non-diverging behavior for $\heff$ in the
dynamics. In the presence of periodic drive, the drive-induced dissipation (DID), drive induced shifts (DIS) 
originated from the second-order terms \cite{chanda2020,chatterjee2020b}. Using FRQME, the second-order
effect of dipolar interaction in the `magic-angle spinning' (MAS) experiment of NMR spectroscopy has also been studied theoretically by this same authors \cite{saha2022}.

\section{Complete set of equations in terms of the observables }
\label{alleqs}
The dynamical equation can be written using 15 observables. The equations can be arranged in a block-diagonal from. Equations of each block is presented here. The general form is given by,
\begin{eqnarray}\label{main}
\dot{A} = \mathbb{L}A + \mathbb{B}
\end{eqnarray}
$A$ is the column vector of the observables, $\mathcal{L}$ is the dynamical matrix, $\mathbb{B}$ is the inhomogeneous term.

The first block consist of $A^{T}_1 = [ M_z \quad M_{zz} \quad M_c]$. the form of $\mathbb{L}$ is given by,
\begin{equation}\label{1eq}
\mathbb{L}_1 = \begin{bmatrix}
-2J-\kappa_1 - 4 \kappa_2 & 0 & 4M_{\circ}\alpha J \\   
M_{\circ}J& -4J-2 \kappa_1 & 2 \alpha J +\kappa_1\\ 
-M_{\circ} \alpha J & 4 \alpha J+2 \kappa_1 & -2J-\kappa_1 \end{bmatrix}    
\end{equation}  
and $\mathbb{B}^T = [2M_{\circ} J  \quad 0 \quad 0]$. the second block is consist of $A^{T}_2 = [ M_x \quad M_{y} \quad M_{xz} \quad M_{yz}]$
\begin{equation}\label{2eq}
 \scriptstyle\mathbb{L}_2  = \begin{bmatrix}
 \scriptstyle-(\frac{5}{2} \kappa_1 +\kappa_2 + 9\kappa_0 +J ) & \scriptstyle-\frac{\delta
\kappa_1}{2}-\delta \kappa_2 +\delta \omega &\scriptstyle -2M_{\circ}J  \alpha &\scriptstyle 2
M_{\circ}\alpha \delta \omega + 6 \omega_{d,0} \\ \scriptstyle  +\frac{\delta \kappa_1}{2}+\delta \kappa_2
-\delta \omega &\scriptstyle -(\frac{5}{2} \kappa_1 +\kappa_2 + 9\kappa_0 +J ) &\scriptstyle -(2
M_{\circ}\alpha \delta \omega + 6 \omega_{d,0})&\scriptstyle-2M_{\circ}J  \alpha\\ \scriptstyle
M_{\circ}J  + \frac{M_{\circ}J  \alpha}{2}& \scriptstyle \frac{M_{\circ} \alpha \delta
\omega}{2}+\frac{3}{2} \omega_{d,0} &\scriptstyle -(\frac{\kappa_1}{2}   +\kappa_2 + 9\kappa_0 + 3J  +2J
\alpha)& \scriptstyle  -\frac{\delta \kappa_1}{2} - \delta \kappa_2 + \delta \omega \\ \scriptstyle
-(\frac{M_{\circ} \alpha \delta \omega}{2}+\frac{3}{2} \omega_{d,0} )&\scriptstyle M_{\circ}J  +
\frac{M_{\circ}J  \alpha}{2}&  \scriptstyle  \frac{\delta \kappa_1}{2} + \delta \kappa_2 - \delta \omega
&\scriptstyle-(\frac{\kappa_1}{2}   +\kappa_2 + 9\kappa_0 + 3J  +2J  \alpha)
\end{bmatrix}    
\end{equation} 
$\mathbb{B}_2$ is a null column vector.
The third block consists of $A^T_3 = [M_{xy}\quad A_c]$. The form of $\mathbb{L}_3$ is given by,
\begin{equation}\label{3eqB}
\mathbb{L}_3 = \begin{bmatrix} -(\kappa_1 + 2\kappa_2 + 2J ) & (\delta \kappa_1 + 2 \delta \kappa_2 - 2 \delta \omega)\\-(\delta \kappa_1 + 2 \delta \kappa_2 - 2 \delta \omega)&
-(\kappa_1 + 2\kappa_2 + 2J )   
\end{bmatrix}
\end{equation} 
$\mathbb{B}_3$ is a null column vector. The fourth block consists of $A^T_4 = [A_{xy}\quad A_z]$.  The form of $\mathbb{L}_4$ is given by,
\begin{equation}\label{4eq}
\mathbb{L}_4 = \begin{bmatrix} -(\kappa_1 + 4 \kappa_0 + 2J ) & ( M_{\circ}\alpha \delta \omega +
\omega_{d,0}  )\\-4( M_{\circ}\alpha \delta \omega + \omega_{d,0}  )&
-(\kappa_1 + 4\kappa_0 + 2J )   
\end{bmatrix}
\end{equation} 
$\mathbb{B}_4$ is a null column vector.The fifth block consists of
$A^{T}_5 = [ A_x \quad A_{y} \quad A_{xz} \quad A_{yz}]$.  The form of $\mathbb{L}_5$ is given by,
\begin{equation}\label{5eq}
 \scriptstyle\mathbb{L}_5  = \begin{bmatrix}
 \scriptstyle-(\frac{1}{2} \kappa_1 +\kappa_2 +  \kappa_0 +J ) & \scriptstyle-\frac{\delta
\kappa_1}{2}-\delta \kappa_2 +\delta \omega &\scriptstyle 2M_{\circ}J  \alpha &\scriptstyle -2
M_{\circ}\alpha \delta \omega + 2 \omega_{d,0} \\ \scriptstyle  +\frac{\delta \kappa_1}{2}+\delta \kappa_2
-\delta \omega &\scriptstyle -(\frac{1}{2} \kappa_1 +\kappa_2 + \kappa_0 +J ) &\scriptstyle 2
M_{\circ}\alpha \delta \omega - 2\omega_{d,0} &\scriptstyle 2M_{\circ}J  \alpha\\ \scriptstyle
M_{\circ}J  - \frac{M_{\circ}J  \alpha}{2}& \scriptstyle -\frac{M_{\circ} \alpha \delta
\omega}{2}+\frac{1}{2} \omega_{d,0} &\scriptstyle -(\frac{\kappa_1}{2}   +\kappa_2 + \kappa_0 + 3J  -2J
\alpha)& \scriptstyle  -\frac{\delta \kappa_1}{2} - \delta \kappa_2 + \delta \omega \\ \scriptstyle
\frac{M_{\circ} \alpha \delta \omega}{2} -\frac{1}{2} \omega_{d,0} & \scriptstyle M_{\circ}J  -
\frac{M_{\circ}J  \alpha}{2}&  \scriptstyle  \frac{\delta \kappa_1}{2} + \delta \kappa_2 - \delta \omega
&\scriptstyle-(\frac{\kappa_1}{2}   +\kappa_2 + \kappa_0 + 3J  -2J  \alpha)
\end{bmatrix}    
\end{equation} 
$\mathbb{B}_5$ is also null column vector.

\bibliographystyle{apsrev4-1}
\bibliography{references1}

\end{document}